%% file: activesteadystate_effectiveT.tex
\documentclass[a4paper,twocolumn,floatfix]{revtex4-1}

\usepackage{bm,amsmath}
\usepackage{amssymb}
\usepackage{graphicx}
\usepackage{hyperref}
\usepackage{xcolor}
\hypersetup{
 colorlinks,
  linkcolor={red!90!yellow},
  citecolor={blue!100!red},
  urlcolor={blue!90!black}
}

\renewcommand{\d}{\mathrm{d}}

\renewcommand{\l}{\lambda}
\newcommand{\p}{\partial}
\newcommand{\f}{\frac}
\renewcommand{\S}{\Sigma}
\renewcommand{\t}{\tau}

\renewcommand{\l}{\lambda}

\newcommand{\DD}{\Delta}
\newcommand{\Tsp}{T_{eff}^{sp}}

{\begin{document}

\title{Viscosity and effective temperature of an active dense system of self-propelled particles}
\author{Saroj Kumar Nandi}
\email{saroj.nandi@weizmann.ac.il}
\affiliation{Department of Chemical and Biological Physics, Weizmann Institute of Science, Rehovot - 7610001, Israel}
 \begin{abstract}
We obtain a nonequilibrium theory for a simple model of a generic class of active dense systems consisting of self-propelled particles with a self-propulsion force, $f_0$, and persistence time, $\t_p$, of their motion. We consider two models of activity and find the system is characterized by an evolving effective temperature $T_{eff}(\t)$, defined through a generalized fluctuation-dissipation theorem. $T_{eff}(\t)$ is equal to the equilibrium temperature at very short time $\t$ and saturates to $T_{eff}=T_{eff}(\t\to\infty)$ at long times; The transition time $t_{trans}$ when $T_{eff}(\t)$ goes to the long-time limit depends on $\t_p$ alone and $t_{trans}\sim \t_p^{0.85}$ for both models. $f_0$ reduces the viscosity with increasing activity, $\t_p$ on the other hand, may increase or decrease viscosity depending on the details of how the activity is included. However, as a function of $T_{eff}$, viscosity shows the same behavior for different models of activity and $\eta\sim (T_{eff}-T)^{-\gamma}$ with $\gamma=1.74$. 
Our theory gives reasonable agreement when compared with experimental data and is consistent with several experiments on diverse systems.

\end{abstract}

\maketitle

\section{Introduction}
Active matter is by definition out of equilibrium system consisting of particles who consume energy and do some work. This broad definition includes a large class of systems, both biological, such as motile cells in tissues \cite{angelini2011,wyart2012,garcia2015} and intracellular cytoplasm \cite{prost2015}, as well as synthetic materials such as Janus particles and light activated swimmers \cite{howse2007,palacci2010,jiang2010,palacci2013} and vertically vibrated granular systems \cite{dauchot2005,nitin2014}. Properties of such systems in the dilute regime have been subjected to extensive investigation in the last couple of decades or so \cite{sriramreview,sriramrmp,catesreview}, however, their study in the dense regime is relatively new. Examples of dense active systems are abundant, starting from cellular cytoskeleton to confluent cells in a tissue. A number of recent simulations \cite{mandal2016,flenner2016,berthier2014,garcia2015,ni2013} and experiments \cite{zhou2009,bechinger2016,angelini2011} on such systems reveal a remarkable similarity with the properties of a glassy system \cite{giulioreview,das2004}. In this work, we consider a simple model of active systems that consists of self-propelled particles (SPP) with a self-propulsion force $f_0$ and persistence time $\tau_p$ of their motion. Such a system, despite its simplicity, comprises important realizations of biological systems and show rich phenomenology through the interplay of activity, order and flow \cite{sriramreview,sriramrmp,catesreview,prost2015,hatwalne2004}. Thus, it is important to extend glass transition theories of passive systems to an active system \cite{berthier2013,saroj2017,sarojPNAS,liluashvili2017,feng2017,szamel2016}. Instead of concentrating on the glass transition itself, here we focus on analyzing two important quantities, the viscosity, which characterizes the rheological properties of a system, and the effective temperature, which plays a crucial role in the description of nonequilibrium systems, within a mean-field theoretical framework applicable to a large number of biological systems.

An important characteristic of active systems is the ability to self-regulate their transport coefficients, such as viscosity, affecting the rheology of the system. For example, the viscosity of cellular-cytoskeleton is much larger in the interphase compared to mitosis when it is dividing \cite{cellbook}. Activity is known to reduce the viscosity of the cell-cortex and {\em in-vitro} assemblies of actin and myosin molecules \cite{humphrey2002,sanchez2012,chan2015}. The behavior of viscosity in active SPP systems is well-understood in the dilute limit \cite{hatwalne2004,sokolov2009,rafai2010,saintillan2010,patterson2016}. Depending on the type of active particles, whether they are contractile or extensile, pullers or pushers, effective viscosity either increases or decreases \cite{hatwalne2004,sokolov2009,rafai2010}. However, how different types of particles in an active dense system affect its viscosity remains an open question. It is particularly important in the biological context since mechanical properties are known to affect functions \cite{samreview}.

Active systems, as they consume energy and perform mechanical work, are by definition out of equilibrium \cite{catesreview,fodor2016}. 
Compared to equilibrium statistical physics, the field of nonequilibrium physics still remains not so well-understood and largely open despite the construction of a number of exact results applicable for such systems in recent times \cite{derrida2007,kurchan2007}.
The search for a description of nonequilibrium systems through an effective temperature ($T_{eff}$) has a long history \cite{hohenberg1989,cugliandolo1997,cugliandolo2011}. A number of recent studies have looked at the behavior of $T_{eff}$ in active systems \cite{shen2004,lu2006,loi2011,loi2011a,wang2011,wang2013,szamel2014}. In driven dense systems, such as a system under a steady shear, $T_{eff}$ is an evolving function of time and its long-time value characterizes the dynamics of the system \cite{berthier2000}. Similarly, an evolving $T_{eff}$ also exists for active SPP systems \cite{saroj2017}, however, the importance of the long-time limit of this variable in the description of different active systems is not yet completely understood.

Our aim in this work is to understand the specific role of activity on the rheological properties and effective temperature of a biological system. Instead of concentrating on the quantitative details of a particular system, we start with a minimal model of an active system and investigate the effect of activity in a broad class of systems. We consider two different models of active SPP systems and obtain a coarse-grained hydrodynamic theory for such systems in their dense regimes. We find that the systems are characterized by an evolving effective temperature $T_{eff}(\t)$, defined through a generalized fluctuation-dissipation relation (FDR), depicting the nonequilibrium nature of the system. $T_{eff}(\t)$ is equal to the equilibrium temperature $T$ at a very short time and saturates to a larger value, $T_{eff}\equiv T_{eff}(\t\to\infty)$, given by the activity parameters, at long times. The transition from $T$ to $T_{eff}$ takes place at a time $t_{trans}$ where $t_{trans}\sim \t_p^{0.85}$ for both models. We obtain the viscosity, $\eta$, through a time-integration of the two-point correlation function. The important findings of this work are as follows: (1) $\eta$ decreases as self-propulsion force, $f_0$, increases and behaves as $\eta\sim f_0^{-3.5}$ for both models. (2) The behavior of $\eta$ as a function of $\t_p$ is more subtle and depends on the details of how the activity is included; it may either decrease or increase with $\t_p$ depending on the active noise statistics for the two models we consider here. (3) $\eta$ as a function of $T_{eff}$ shows the same behavior for both systems, and both parameters $f_0$ and $\t_p$, revealing that $T_{eff}$ works as a rational control parameter for such systems in their dense regime. Comparison of our theory with existing experimental data show reasonable agreements and we conclude by discussing ways to test our theory in experiments on biological systems.

\section{Theory}
We start with the equation of motion for the density fluctuation $\phi(t)$ at time $t$ of a generic class of active dense fluid. The active energy in the system is continuously supplied through the microscopic degrees of freedom. This energy can be supplied externally, as in the case of vibrated granular particles \cite{dauchot2005,nitin2014}, or it may come from the solvent through ATP (adenosine tri-phosphate) as in biological systems \cite{angelini2011,zhou2009}. We characterize such active driving through a colored noise and write the Langevin equation for $\phi(t)$ as
\begin{equation}
\label{langevin}
 \f{\p\phi(t)}{\p t}+\mu(t)\phi(t)=-\f{g}{2}\phi^2(t)+\xi(t)+f(t)
\end{equation}
where $\mu(t)$ is a frequency term and $g$ encodes the interactions in the fluid. $\xi(t)$ is the thermal noise with zero mean and $\langle\xi(t)\xi(t')\rangle=2T\delta(t-t')$ with $T$ being the ambient temperature. $f(t)$ is the active noise with zero mean and correlation $\langle f(t)f(t')\rangle=2\Delta(t-t')$. We have set $k_B$, the Boltzmann constant, to unity. We have ignored frictional dissipation for simplicity, as friction merely sets a length scale. Our model can be seen as the schematic form of the equation of motion for density fluctuation at a particular wave vector, corresponding to the first maximum of static structure factor, for an active fluid obtained from the continuity equations for density and momentum density \cite{saroj2017}. Eq. (\ref{langevin}) is a minimal model for active systems of self-propelled particles \cite{prost2009,benisaac2011,benisaac2015,henkes2011,bohec2013,berthier2013}.

In equilibrium, one obtains the two-point correlation function starting from Eq. (\ref{langevin}) [in the absence of the active noise] and this characterizes the dynamics entirely since correlation and response functions are related via FDR. However, there is no such simple relation in nonequilibrium systems and one must look at the equations of motion for both the correlation and response functions. We use the field theoretic method of mode-coupling theory (MCT) \cite{reichman2005,castellani2005,kuni2005,kim2008,saroj2017,saroj2012,saroj2016} for our calculation. We first obtain the equations of motion for the correlation, $C(t,t')=\langle \phi(t)\phi(t')\rangle$, and the response, $R(t,t')=\langle \p\phi(t)/\p \xi(t')\rangle$, functions for a generic nonequilibrium non-stationary state of the active system as
\begin{align}\label{nonst1}
 \f{\p C(t,t')}{\p t}=&-\mu(t)C(t,t')+\int_0^{t'}\d sD(t,s)R(t',s) \nonumber\\
 +&\int_0^t\d s\S(t,s)C(s,t')+2TR(t',t) \\
 \f{\p R(t,t')}{\p t}=&-\mu(t)R(t,t')+\int_{t'}^t\d s\S(t,s)R(s,t')+\delta(t-s) \label{nonst2}\\
 \mu(t)=T+&\int_0^{t}\d s[D(t,s)R(t,s)+\S(t,s)C(t,s)]\label{nonst3}
\end{align}
with $D(t,s)=2\lambda C^2(t,s)+\Delta(t-s)$ and $\S(t,s)=4\l C(t,s)R(t,s)$ where, following standard notation, we have used $g^2=4\l$. Similar equations were also obtained in \cite{berthier2013} for a $p$-spin spherical active spin-glass model as well as in \cite{saroj2017} for a hydrodynamic model of active fluid. It is advantageous to use an integrated response function, $F(t,t')=-\int_{t'}^tR(t,s)\d s$, since the behavior of $F(t,t')$ is smoother compared to that of $R(t,t')$. We now assume that the system goes to a steady state at long time and the two-time functions become functions of the time-difference only: $C(t,t')=C(t-t')=C(\t)$ and $F(t,t')=F(t-t')=F(\t)$, in the steady state. Then, after a straightforward but tedious algebra, we obtain the equations of motion for the correlation and integrated-response functions describing the steady state of the active system as
\begin{align} \label{activemct1}
 \f{\p C(\t)}{\p \t}&=\Pi(\t)-(T-p)C(\t)-\int_0^\t m(\t-s)\f{\p C(s)}{\p s}\d s \\
 \f{\p F(\t)}{\p \t}&=-1-(T-p)F(\t)-\int_0^\t m(\t-s)\f{\p F(s)}{\p s}\d s \label{activemct2}
\end{align}
\begin{align}
\text{where, } \hspace{0.1cm}& m(\t)=2\l\f{C^2(\t)}{T_{eff}(\t)}; \,\,\,
 p=\int_0^\infty \DD(s)\f{\p F(s)}{\p s}\d s \label{modelp}\\
\text{and } &\Pi(\t)=-\int_\t^\infty \DD(s)\f{\p F(s-\t)}{\p s}\d s \label{modelPi},
\end{align}
and $T_{eff}(\t)$ is defined through a generalized FDR as
\begin{equation}
\label{teff}
 \f{\p C(\tau)}{\p \tau}=T_{eff}(\t)\f{\p F(\tau)}{\p \tau}.
\end{equation}
We are interested in the glass transition of the system in this work, instead, we concentrate on the behavior of viscosity and $T_{eff}(\t)$ as functions of the control parameters, $f_0$ and $\t_p$, when the system is still a liquid.
Eqs. (\ref{activemct1}-\ref{teff}) provide the MCT equations of motion for the correlation and integrated-response functions in steady-state of the active system defined in Eq. (\ref{langevin}). Note that the same set of equations were obtained in \cite{saroj2017} starting with the continuity equations of density and momentum density for an active system.
To complete the description we must provide the active noise statistics, $\DD(\t)$, appearing in Eqs. (\ref{modelp}) and (\ref{modelPi}) above. Mainly two types of statistics have been used in the literature \cite{mandal2016,flenner2016,paoluzzi2018,benisaac2015} as has been discussed in Ref. \cite{saroj2017}: 
\begin{itemize}
 \item SNTC (shot noise temporal correlation) noise statistics defined as $\DD(\t)=\DD_0\exp(-\t/\t_p)$.
 \item OUP (Ornstein-Uhlenbeck process) noise statistics, $\DD(\t)=(\Tsp/\t)\exp(-\t/\t_p)$.
\end{itemize}
$\DD_0$ and $\Tsp$ are proportional to $f_0^2$. The system is driven away from the glassy regime as a function of $\DD_0$ or $\Tsp$ within both the noise statistics. However, $\t_p$ inhibits glassiness within SNTC statistics whereas it promotes glassiness within OUP statistics as we show below \cite{saroj2017}. 

\begin{figure}
 \includegraphics[width=8.6cm]{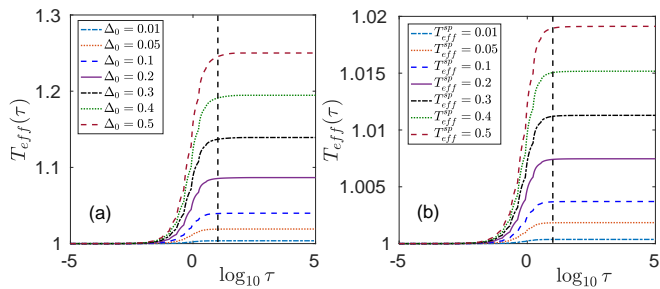}
 \caption{Evolution of $T_{eff}(\t)$ defined through a generalized fluctuation-dissipation relation. (a) $T_{eff}(\t)$ as a function of $\t$ within SNTC statistics for $\t_p=10.0$ and different $\DD_0$ as shown in the figure. (b) $T_{eff}(\t)$ as a function of $\t$ within OUP statistics with $\t_p=10.0$ and different $\Tsp$ as shown in the figure. The transition of $T_{eff}(\t)$ from $T$ to $T_{eff}$ takes place at similar time for all the curves showing that this transition is determined by $\t_p$ (see Fig. \ref{teff_bothmodels_fixedf0}) that is kept fixed in both the figures.}
 \label{teff_fixedtaup}
\end{figure}

\section{Results}
Analytic solutions of Eqs. (\ref{activemct1}-\ref{modelPi}) in general is not possible and they must be solved numerically; $T_{eff}(\t)$, as defined by Eq. (\ref{teff}), needs to be evaluated at each time-step. We fix $T=1.0$ and present the results in terms of $\l$. The solutions of these equations are well-known in the absence of activity \cite{goetzebook,das2004,leutheusser1984}. $C(\t)$ decays rapidly at small $\l$ (or large $T$) and develops a two-step relaxation scenario when $\l$ becomes close to but smaller than $2.0$; $C(\t)$ first rapidly decays to a plateau and then shows a much slower decay from the plateau to zero at long times. When $\l=2.0$ and beyond, $C(\t)$ no longer decays to zero; this is the well-known nonergodicity transition within MCT \cite{goetzebook,das2004}. In this work, we restrict ourselves in the ergodic phase since the non-ergodic phase is not physically relevant.
We first look at the general behavior of $T_{eff}(\t)$ as a function of self-propulsion. $T_{eff}(\t)$ has an evolving nature within both the noise statistics as shown in Figs. \ref{teff_fixedtaup}(a) and (b) as a function of $\t$ with $T=1.0$, $\l=2.0$ and $\t_p=10.0$; $T_{eff}(\t)$ is equal to $T$ at short time and evolves to a different value, $T_{eff}=T_{eff}(\t\to\infty)$, at long time. The nonequilibrium nature of the system is manifested through this time-dependent $T_{eff}(\t)$ that saturates roughly at the same time for all $\DD_0$ or $\Tsp$ within both statistics and this saturation time is of the order of $\t_p$; the value of $\t_p$ is shown by the vertical dotted lines in both figures. Analytical expressions for the dependence of $T_{eff}$ on $f_0$ and $\t_p$ have been presented in \cite{saroj2017}.

\begin{figure}
 \includegraphics[width=8.6cm]{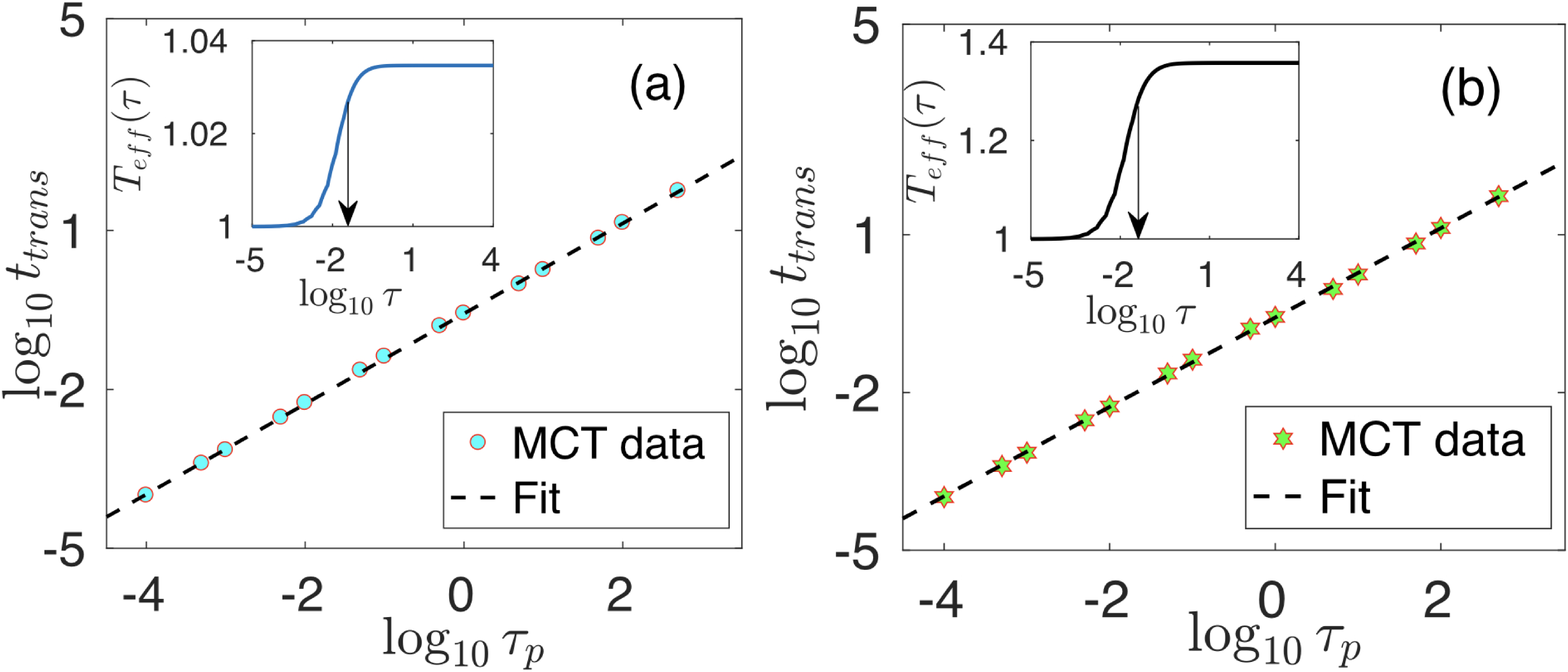}
 \caption{The time when $T_{eff}(\t)$ goes from $T$ to $T_{eff}$ is denoted as $t_{trans}$. Since $T_{eff}(\t)$ evolves continuously, in practice we define $t_{trans}$ as the time when $(T_{eff}(\t)-T)$ becomes $80\%$ of its final value. We show the behavior of $t_{trans}$ as a function of $\t_p$ for SNTC statistics in (a) with $\DD=0.5$ and for OUP statistics in (b) where we have used $\Tsp=0.5$. In the insets of both figures, we show the behavior of $T_{eff}(\t)$ as a function of $\t$ with $\t_p=0.1$. The arrows show the values of $t_{trans}$. }
 \label{teff_bothmodels_fixedf0}
\end{figure}

It is expected that $\t_p$ somehow be related to the evolution of $T_{eff}$ since $\t_p$ controls the temporal correlations of the noise statistics. To understand this evolution of $T_{eff}(\t)$ we now keep $\DD_0$ or $\Tsp$ fixed (depending on the model), and look at the transition of $T_{eff}(\t)$ from $T$ to $T_{eff}$. We define the transition time $t_{trans}$ as the time when $T_{eff}(\t)-T$ becomes $80\%$ of its final value as shown in the insets of Fig. \ref{teff_bothmodels_fixedf0}. This definition is somewhat arbitrary, however, it helps to calculate $t_{trans}$ from the numerical solution of Eqs. (\ref{activemct1}-\ref{teff}) since the approach to the final value is quite slow and it is hard to obtain the exact time when $T_{eff}(\t)$ reaches its long-time value. 
We have checked that other definitions lead to the same result.
We show the behavior of $t_{trans}$ as a function of $\t_p$ for SNTC statistics in Fig. \ref{teff_bothmodels_fixedf0}(a) where we have kept $\DD_0=0.5$ fixed and for OUP statistics in Fig. \ref{teff_bothmodels_fixedf0}(b) where we have used $\Tsp=0.5$. In the insets of both these figures, we show the behaviors of $T_{eff}(\t)$ as a function of $\t$. The arrows in these figures indicate the value of $t_{trans}$. Both the fits in Fig. \ref{teff_bothmodels_fixedf0}(a) and (b) show $t_{trans}\sim \t_p^{0.85}$. Thus, we see that in the dense regime $T_{eff}(\t)=T$ when $\t\ll\t_p$ and saturates to $T_{eff}$, determined by the activity parameters when $\t\gg\t_p$; the transition from $T$ to $T_{eff}$ takes place at a time $t_{trans}\sim \t_p^{0.85}$. We have seen that $T_{eff}(\t)$ always has this evolving nature and $T_{eff}(\t)\geq T$ in general. Similar evolving $T_{eff}(\t)$ has also been reported for motorized particles \cite{shen2004} where the behavior is even more complex with a non-monotonic evolution.

\begin{figure}
 \includegraphics[width=8.6cm]{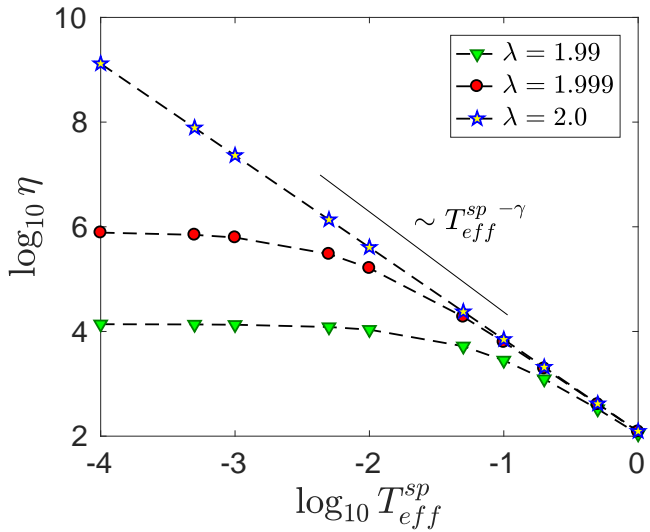}
 \caption{Viscosity $\eta$ as a function of $\Tsp$ (Eq. \ref{viscosity}) within OUP statistics with $\t_p=10.0$, $T=1.0$ and three values of $\l$ as shown in the figure. When $\l<\l_c=2.0$, $\eta$ saturates at small $\Tsp$ and active thinning (reduction in viscosity) is observed at larger $\Tsp$. When $\l=\l_c$ or larger, any amount of activity shows active thinning since viscosity of the passive system diverges within the theory. $\eta\sim {\Tsp}^{-\gamma}$ in the activity dominated regime with $\gamma=1.74$.}
 \label{visc_tsp}
\end{figure}

Next, we look at the behavior of viscosity, $\eta$. In equilibrium, we obtain $\eta$ through Kubo relation by integrating $C(\t)$ for all $\t$ and divided by temperature. In general, we don't have such a relation for nonequilibrium systems. However, considering the departure of the system from equilibrium is small when activity is not very large, such that we are in the linear-response regime \cite{kubo1966}, we can still apply Kubo \footnote{We have also used a different definition: $\eta=\int_0^\infty [C(t)/T_{eff}(t)]\d t$ and find that it leads to similar results as the definition above. The reason is the time $t$, when $T_{eff}(t)$ is significantly different from $T$, $C(t)$ has almost decayed to zero by then.} formula and obtain $\eta$ as
\begin{equation}\label{viscosity}
 \eta=\f{1}{T}\int_0^\infty C(t)\d t.
\end{equation}
Self-propulsion drives the system away from the glassy regime and therefore, we expect the viscosity to decrease as a function of self-propulsion as is shown in Fig. \ref{visc_tsp} for different values of $\l$ for OUP statistics.
When $\l<2.0$, a small amount of activity ($\Tsp$) does not affect the dynamics and viscosity saturates to that of the passive system. However, the viscosity of the passive system becomes $\infty$ for $\l=2.0$ and any amount of activity is going to affect the value of $\eta$. In the regime where activity dominates, we find $\eta\sim {\Tsp}^{-\gamma}$ with $\gamma=1.74$, the same exponent that governs the behavior of passive systems \cite{saroj2017}. The effect of activity in this aspect is similar to that of shear \cite{brader2009,fuchs2002} that cuts-off relaxation and reduces the viscosity \footnote{Note that driving a system through shear and activity has important differences; any non-zero shear rate cuts-off the non-ergodicity transition within MCT whereas an active system may still show this transition at low activity.}. The qualitative behavior of $\eta$ as a function of $\DD_0$ within SNTC statistics is similar. Thus, self-propulsion always show active thinning (in analogy with shear thinning \cite{brader2009}) where activity reduces the viscosity.

\begin{figure}
 \includegraphics[width=8.6cm]{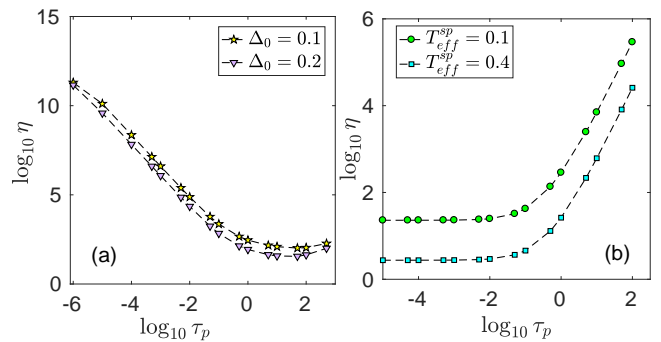}
 \caption{Viscosity, as defined in Eq. (\ref{viscosity}), obtained through the numerical solution of Eqs. (\ref{activemct1}-\ref{teff}) with $T=1.0$ and $\l=2.0$. (a) $\eta$ as a function of $\t_p$ for two different values of $\DD_0$ with SNTC statistics. $\eta$ decreases with increasing $\t_p$ showing active thinning within this noise statistics. (b) $\eta$ as a function of $\t_p$ within OUP statistics for two different $\Tsp$ as shown in the figure. $\eta$ within this noise statistics increases with activity showing active thickening.}
 \label{visctaup}
\end{figure}

The effect of $\t_p$ on the viscosity of the system is more subtle and strongly depends on the microscopic details of how activity is implemented. The behavior within both the statistics are shown in Fig. \ref{visctaup}. We have kept $T=1.0$ and $\l=2.0$ fixed for this figure. $\t_p$ within the SNTC noise statistics drives the system away from glassy regime \cite{saroj2017,mandal2016} and the viscosity decreases as $\t_p$ increases as shown in Fig. \ref{visctaup}(a) for two values of $\DD_0$. On the other hand, $\t_p$ within the OUP noise statistics drives the system towards the glassy regime \cite{saroj2017,flenner2016} and the viscosity increases with increasing $\t_p$. Thus, $\t_p$ shows contrasting behavior within the two models: activity, controlled through $\t_p$ shows active thinning within SNTC statistics whereas it shows active thickening within OUP statistics. These results are consistent with the behavior of relaxation time within the two statistics \cite{saroj2017,sarojPNAS}. We discuss ways to test these predictions below.

\begin{figure}
 \includegraphics[width=8.6cm]{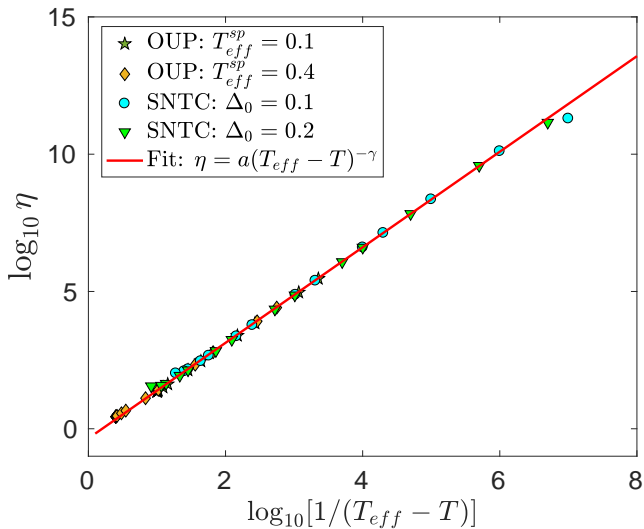}
 \caption{Mode-coupling theory predicts $\eta\sim (T_{eff}-T)^{-\gamma}$ with $\gamma=1.74$ for an active system. We test this by plotting the viscosity $\eta$, obtained for the two noise statistics that we consider for different $\DD_0$, $\Tsp$ and $\t_p$ as a function of $T_{eff}$ and find that they follow a master curve that fits well with the MCT prediction. This shows the possibility of $T_{eff}$ being a rational parameter for understanding the dynamics of an active dense system of self-propelled particles.}
 \label{viscTeff}
\end{figure}

We now obtain $\eta$ as a function of $T_{eff}$ for both models. 
Mode-coupling theory for passive systems predicts a divergence of viscosity at the MCT transition: $\eta\sim (\sigma-\sigma_c)^{-\gamma}$, where $\sigma$ is the control parameter ($T$ or $\l$) and $\sigma_c$ denotes the MCT transition point \cite{goetze1987,goetze1988}. Then, for the active system, with the values of parameters such that the passive system is at the transition, we obtain $\eta\sim (T_{eff}-T)^{-\gamma}$. This is an important prediction, readily testable in experiments and simulations, of the theory, which says that irrespective of how $\eta$ changes as functions of the control parameters for different models, the behavior of $\eta$ as a function of $T_{eff}$ is same. We plot the viscosity for different $\DD_0$ and $\Tsp$ for the two models as presented in Fig. \ref{visctaup} as function $T_{eff}$ in Fig. \ref{viscTeff} and find that the data follow a master curve. We fit the data with $\eta=a(T_{eff}-T)^{-\gamma}$ and obtain $a=0.45$ and $\gamma=1.74$. The value of $\gamma$ is same as that for a passive system corroborating the theoretical prediction. This shows the possibility of $T_{eff}$ being an important parameter to understand the dynamics of active SPP systems in their dense regime.

\begin{figure}
 \includegraphics[width=8.6cm]{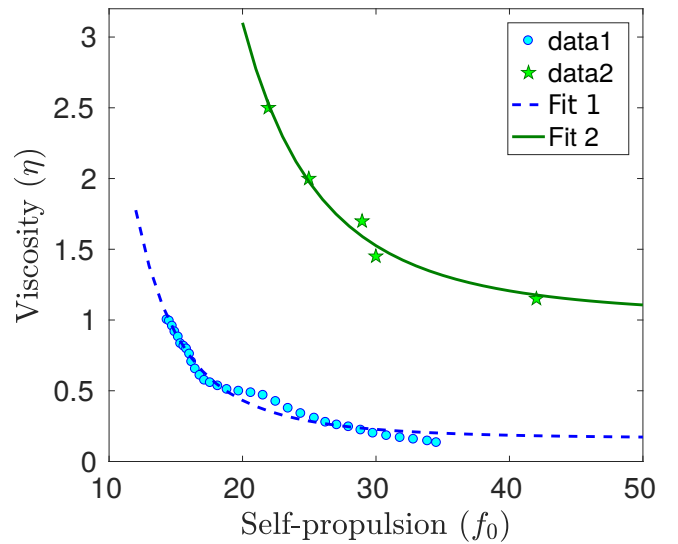}
 \caption{Data 1 and Data 2 are obtained from Ref. \cite{sokolov2009} and \cite{rafai2010} respectively for viscosity as function of self-propulsion velocity (which is proportional to $f_0$). The lines are fits of our theoretical prediction $\eta=a+bf_0^{-3.5}$ with   $a=0.16$ and $b=9.7\times 10^3$ for fit 1 and $a=1.02$ and $b=7.4\times 10^4$ for fit 2.}
 \label{comp_expt}
\end{figure}

Let us now compare this prediction of the theory with available experimental data. In terms of self-propulsion force $f_0$, the theory predicts $\eta\sim f_0^{-3.5}$. Refs. \cite{rafai2010,sokolov2009} have looked at the viscosities of bacteria and motile microalgae suspensions. Although the experiments were performed with an interest in the dilute regime, the boundary dividing the two regimes of dilute and dense is not sharp. Moreover, a close look at the mean-square displacement, as in Fig. 1(b) of Ref. \cite{rafai2010}, shows a similarity to that of a dense fluid.
We have collected the data of $\eta$ as a function of self-propulsion velocity (that is proportional to $f_0$) from Fig. 1(c) of Ref. \cite{rafai2010} and from Fig. 5, corresponding to the larger density $n=1.8\times 10^{10} cm^{-3}$, of Ref. \cite{sokolov2009} and present the data in Fig. \ref{comp_expt}. The lines are fits to our theoretical prediction, $\eta=a+b f_0^{-3.5}$. Both the sets of data seem to agree reasonably well with the prediction. It will be desirable to have systematic experimental data in more dense suspensions for further tests of the theory. Our theory is also consistent with the increase in viscosity with ATP depletion in an amphibian oocyte nucleolus \cite{brangwynne2011}.

\section{Discussion}
We have obtained a coarse-grained hydrodynamic theory for the dense regime of two general models of active matter systems consisting of self-propelled particles with a self-propulsion force, $f_0$, and persistence time, $\t_p$, of their motion. In this work, we have concentrated on the behavior of two important quantities, namely the fluid viscosity, $\eta$, that governs the rheological properties and the effective temperature, $T_{eff}(\t)$, which is important for the description of nonequilibrium systems.
We find an evolving $T_{eff}(\t)$ that is equal to the equilibrium temperature at a short time and saturates to $T_{eff}=T_{eff}(\t\to\infty)$ at long times. The transition from $T$ to $T_{eff}$ takes place at a time $t_{trans}$ that is given by the persistence time of the particles; $t_{trans}\sim \t_p^{0.85}$. Since $\t_p$ controls the temporal correlations of the active noise statistics, it is expected that $t_{trans}$ is related to $\t_p$. The evolving nature of $T_{eff}(\t)$ is quite similar to that of another class of driven systems, glassy systems under shear \cite{ono2002,haxton2007,berthier2000}, where $t_{trans}$ is determined by the shear rate. 

The viscosity $\eta$ decreases as $f_0$ at a fixed $\t_p$ increases showing active thinning behavior. However, when activity is controlled through $\t_p$ at fixed $f_0$, the behavior of $\eta$ depends on details of how the activity is included. Within the two active noise statistics that we consider here, $\t_p$ with SNTC statistics shows active thinning whereas $\t_p$ within OUP statistics shows active thickening where viscosity increases with activity. The theory predicts $\eta\sim(T_{eff}-T)^{-\gamma}$ independent of the model and the numerical solution of Eqs. (\ref{activemct1}-\ref{teff}) supports this prediction. This is an important generic prediction of the theory easily verifiable in experiments. $(T_eff-T)$ is proportional to $f_0^2\t_p/(1+A\t_p)$ and $f_0^2/(1+A\t_p)$, where $A$ is a constant for SNTC and OUP statistics respectively \cite{saroj2017}. This shows the possibility of $T_{eff}$ being a rational parameter for the dynamics of active dense systems. Comparison of the prediction with available experimental data (Fig. \ref{comp_expt}) show reasonable agreement.

We have taken a minimal model, Eq (\ref{langevin}), of active matter systems as our starting point as we are interested in a broad understanding of the effect of activity on diverse biological systems. Eq. (\ref{langevin}) can be viewed as a simplified version of a more complicated system. Including elements of greater detail are not going to affect our qualitative results. Memory, in the form of a generalized relaxation kernel, as in Eqs. (\ref{activemct1}-\ref{modelp}), are indeed quite common in the dynamics of biological systems, both for intracellular  \cite{frank2016} as well as inter-cellular \cite{selmeczi2005} dynamics, and requires more detailed exploration. It will be interesting to test our theory in dense suspensions of active particles, like bacteria and algea as in \cite{sokolov2009,rafai2010}. The SNTC noise statistics is relevant for intracellular dynamics as in \cite{zhou2009,deng2006,bursac2005} as well as {\em in-vitro} assemblies of actomyosin \cite{humphrey2002} or that of microtubule and kinesin molecules \cite{sanchez2012}. On the other hand, OUP noise statistics is relevant for systems like a cellular monolayer \cite{angelini2011,sepulveda2013,das2015} or a collection of active particles in suspension \cite{sokolov2009,rafai2010}. Our theory predicts the same behavior of $\eta$ as a function of $f_0$, however, the predictions as function of $\t_p$ are quite distinct for the two classes of systems and, thus, easily testable in experiments. For example, changing the merlin proteins through RNAi experiments in a monolayer of cells (e.g., Madin-Darby canine kidney cells) \cite{das2015} gives one way of changing the persistence time in such a system. We look forward to further tests of our theory.

\section{Acknowledgements}
We thank L. Berthier, M. Cates, C. Dasgupta, O. Dauchot, N. S. Gov, Frank J{\"{u}}licher, Smarajit Karmakar, J. Kurchan, J. Prost,  M. Rao, S. Ramaswamy, Thomas Speck, G. Szamel and Thomas Voigtmann for many important discussions and Koshland foundation for funding through a fellowship.

\input{activesteadystate_effectiveT.bbl}


\end{document}

%% file: activesteadystate_effectiveT.bbl
%